\documentclass[aps, prfluids,superscriptaddress,longbibliography]{revtex4-2}

\usepackage{graphicx}
\usepackage{dcolumn}
\usepackage{bm}
\usepackage{booktabs}
\usepackage{setspace}
\usepackage{amsmath}
\usepackage{mathtools}
\usepackage{amssymb}
\usepackage{chngcntr}

\graphicspath{{./figures/}}

 \newcommand{\ub}{\mathbf{u}}
 \newcommand{\vb}{\mathbf{v}}
 
 \newcommand{\xb}{\mathbf{x}}

 \newcommand{\St}{\mathrm{St}}
 \newcommand{\re}{\mathrm{Re}} 
 \newcommand{\eps}{\mathrm{\gamma}}

 \newcommand{\Kbp}{\mathbf{\hat{K}}}
 
 \newcommand{\Kbpp}{\mathbf{\hat{\hat{K}}} }

\begin{document}

\title{Anisotropic Particles Focusing Effect in Complex Flows}
\author{S\'everine Atis}
\affiliation{Massachusetts Institute of Technology, Department of Mechanical Engineering, Cambridge, Massachusetts 02139.}
\affiliation{University of Chicago, Department of Physics, Chicago, Illinois 60637.}
\author{Matthieu Leclair}
\affiliation{Massachusetts Institute of Technology, Department of Mechanical Engineering, Cambridge, Massachusetts 02139.}
\author{Themistoklis P. Sapsis}
\affiliation{Massachusetts Institute of Technology, Department of Mechanical Engineering, Cambridge, Massachusetts 02139.}
\author{Thomas Peacock}
\affiliation{Massachusetts Institute of Technology, Department of Mechanical Engineering, Cambridge, Massachusetts 02139.}

\begin{abstract}
Finite size effects can lead neutrally buoyant particles to exhibit different dynamics than tracer particles, and can alter their transport properties in fluid flows. Here we investigate the effect of the particle's shape on their dispersion in 2-dimensional complex flows. Combining numerical simulations with laboratory experiments, we show that particles with isotropic and anisotropic shapes exhibit different Lagrangian coherent structures, resulting in distinct dispersion phenomena within a given flow field. Experiments with rod shaped particles show a focusing effect in the vicinity of vortex cores. We present a simple model that describes the dynamics of neutrally buoyant ellipsoidal particles in two-dimensional flow and show that particle aspect ratio and orientation dependent forces can generate clustering phenomena in vortices.
\end{abstract}

\maketitle

\today

\section{Introduction}

The dispersion of a tracer in a fluid flow is influenced by the Lagrangian motion of fluid elements. Even in laminar regimes, the irregular chaotic behavior of a fluid flow can lead to effective stirring that rapidly redistributes tracers throughout the domain. In this context, small finite-size objects can behave considerably different from infinitesimal fluid particles, and lead to altered clustering and dispersion phenomena. Ranging from pollutant transport \cite{peacock2013} and plankton population evolution on the surface of the ocean \cite{abraham1998}, to interstellar dust dynamics \cite{gold1952,andersson2015}, 
quantifying the impact of finite-size effects is therefor essential to describe material transport in natural and industrial processes.

However, depending on their size and buoyancy properties, inertial particles can exhibit a variety of dynamics, resulting in markedly different trajectories \cite{haller08, ouellette08, brown09, sudharsan2016}.
The shape of a particle can also affect its response to fluid flow \cite{zhang01, ni14,parsa2012,voth2017}.
For instance, in cloud formation the anisotropic shapes of ice particles influence their transport \cite{ni14,Jucha2018} and aggregation \cite{saw08, squires90} in turbulent atmospheric flows.
Fiber transport in cellular flows, on the other hand, have been shown to exhibit hydrodynamic dispersion phenomena controlled by the buckling properties of the material \cite{wandersman2010, quennouz2015}.
Under marine conditions, preferential alignment of elongated microorganisms in the local flow direction can affect light backscattering and change global carbon fixation rates within phytoplankton blooms \cite{marcos11}.
In recent studies, predictive models suggested that the irregular shape of microplastic contaminants could also have an impact on their long term transport properties on the ocean surface \cite{broday1997,chubarenko2016, dibenedetto2018}.

In this paper, we investigate the effect of anistropic shape on the particles dynamics and long term behavior in chaotic flows.
We experimentally show that when advected in two-dimensional cellular flows for an extended period of time, finite-size particles trajectories with an ellipsoidal shape start diverging from the spherical ones. The ellipsoidal particles exhibit a strong clustering in the vicinity of vortices with an aggregation rate that depends on the particle's aspect ratio.
We present a simple model that combines finite-size effects with an orientational-dependent dynamics, and perform in parallel, numerical simulations under identical conditions to the experiments. 
Our results suggest that neutrally buoyant ellipsoidal particles present stronger inertial effects, and are attracted to the coherent regions of the flow at a higher rate than spherical particles.

\section{Finite-size anisotropic particles dynamics}

\subsection{Isotropic inertial particles}

Inertial effects arise when the advected particle and fluid flow accelerations become different; namely, the particle is responding to changes in the underlying flow field over a finite time and additional forces are generated. For instance when the particle density is different from the carrying fluid density, even for a small difference, these effects can accumulate over time and eventually lead to a drastic deviation between the fluid flow and particle trajectory.
Experiments conducted in 3-dimensional turbulent flows \cite{laporta01, brown09, volk11, calzavarini12, klein13}, and in 2-dimensional chaotic flows \cite{babiano00, ouellette08}, have shown that the particles behavior can be partly quantified by the particle's Stokes number $\St$:
\begin{equation}
\mathrm{St} = \frac{2}{9}\frac{\rho_p}{\rho_f}\left(\frac{a}{L}\right)^2\mathrm{Re} \text{,}
\end{equation}
where $\rho_p$ is the particle density, $\rho_f$ the fluid density, $a$ the particle radius, $L$ the flow characteristic length scale, and $\re = UL/\nu$ is the flow Reynolds number with $\nu$ the fluid kinematic viscosity.
Assuming $a \ll L$, the dynamics of small but finite-size inertial particles can be described with a simplified  Maxey-Riley equation \cite{maxey83}, neglecting the Fax\'en correction and memory terms \cite{michaelides97, babiano00, haller08, sapsis11, suppmat}.
After rescaling space by $L$, and time by $T=L/U$, with $U$ the characteristic flow velocity, the momentum equation becomes \cite{sapsis11}
\begin{equation}\label{MX1}
\frac{d \vb}{dt} =  \frac{3R}{2}\frac{D\ub}{Dt}  - \frac{R}{\St} (\vb - \ub) + \left ( 1 - \frac{3R}{2}\right ) \mathbf{g} \text{.}
\end{equation}
The particle's velocity $\vb$ depends on three forces in the right hand side of the equation: the first term corresponds to advection and added mass, where $\ub$ is the flow velocity, the second term is the Stokes drag force, and the third term is the buoyancy force, where the density ratio is given by $R=2 \rho_f / (\rho_f + 2 \rho_p)$, and $\mathbf{g}= -g\ \mathbf{e_z}$ is gravity. Note that $d\mathbf{.}/dt = \frac{\partial .}{\partial t} + (\vb \cdot \nabla).$ and $D\mathbf{.}/Dt = \frac{\partial .}{\partial t} + (\ub \cdot \nabla).$ are different here \cite{babiano00}.

When the particles are neutrally buoyant, such that $\rho_p = \rho_f$, $R = 2/3$ and Equation \eqref{MX1} reduces to \cite{sapsis11}
\begin{equation}\label{MX3}
\frac{d \vb}{dt} =  \frac{D\ub}{Dt}  - \frac{2}{3}\St^{-1} (\vb - \ub) \text{.}
\end{equation}
This shows that particles with finite $\St$ number can still exhibit a dynamics with a finite relaxation time due to the flow interactions with the particle's surface \cite{babiano00, haller08, ouellette08, brown09}.
Note that equation \eqref{MX3} converges toward the fluid elements motion when $\St\rightarrow0$ for infinitesimal particles.

\subsection{Anisotropic particles angular dynamics}

In comparison with spherical particles, anisotropic particles can exhibit additional torques and forces due to the coupling between their rotational and translational behavior.
The motion of small, neutrally buoyant ellipsoidal particles in laminar flow has been first studied by G. B. Jeffery in 1922 \cite{jeffery22}. His theoretical work on the torque experienced by non-inertial ellipsoidal particles combined with more recent numerical simulations with orientation dependent relaxation times \cite{zhang01,mortensen08,dibenedetto2018}, suggest that the dynamics of even very small objects, i. e. $\mathrm{St}\ll1$, could be affected by their anisotropic shape. Experiments in turbulent flow, have also indicated that the turbulent kinetic energy of a suspension is altered accordingly with the suspended particles' shape \cite{bellani12}. Finally, numerical models with small inertial effects coupled with orientation dependent forces, showed that the transport of anisotropic particles under a wave can also lead to distinct behaviors over long period of times \cite{dibenedetto2018}.

Neutrally buoyant ellipsoidal particles experience a torque that tends to align their principal axis parallel to the surrounding fluid principal axis of distortion. For non-inertial particles, the particle's velocity is identical to the surrounding fluid $\vb = \ub$ at all time. In this situation, there is no resultant force and the torque acting on the particle vanishes at first order \cite{jeffery22}. The particle angular velocity $\dot{\theta}$ is then simply expressed as a function of the flow vorticity $\omega$ and strain-rate: \cite{olson98, parsa11, suppmat}
\begin{equation}\label{thetadot}
\dot{\theta} = \frac{\omega}{2}  + \frac{1}{2}  \left(\frac{1-\eps^2}{1+\eps^2}\right) \left[ \sin2\theta \left( \frac{\partial u_x}{\partial x}-\frac{\partial u_y}{\partial y} \right) - \cos2\theta \left( \frac{\partial u_y}{\partial x} + \frac{\partial u_x}{\partial y} \right) \right]
\end{equation}
where $\eps=a_2/a_1$ is the particle aspect ratio, with $a_1$ and $a_2$ the semi-minor and semi-major axes respectively, and $u_x$ and $u_y$ are the horizontal and vertical components of the flow velocity in the fixed frame. The particle's orientational dynamics is described using 3 coordinate systems: a fixed frame $\xb = (x,y,z)$, a particle frame $\hat{\xb} = (\hat{x},\hat{y},\hat{z})$, and a co-moving frame $\hat{\hat{\xb}} = (\hat{\hat{x}},\hat{\hat{y}},\hat{\hat{z}})$, see supplemental Fig. C.1 for a schematic \cite{suppmat}.
In this configuration, $\theta$ corresponds to the angle between the particle's major axis and $\mathbf{x}$ or $\mathbf{\hat{\hat{x}}}$.

\subsection{Anisotropic inertial particles}

To describe the combined effects of finite-size and anisotropic shape with a minimal model, we present here an equation of motion that couples the particle's orientation with the drag force only.
In order to consider a finite-size particle, we need to take into account inertial forces that will tend to separate the particle's trajectory from the fluid trajectory, such that now $\vb \neq \ub$. 
The dynamics of anisotropic particles is not well approximated by equation \eqref{MX3} anymore, and need to be modified to account for the orientation specific dynamics \cite{gallily79-2, koch89}. For instance, the added mass and drag force terms can become dependent to the particle's instantaneous orientation \cite{mortensen08, zhang01,dibenedetto2018}.
Here, we use a linear momentum equation with a the hydrodynamic drag force term coupled with the ellipsoidal particle's orientation \cite{brenner64-3, brenner64-4, fan95, loth08}
\begin{equation}
\mathbf{F} = -6 \pi a_{s} \mu \; \Kbpp (\vb- \ub) \text{,}
\end{equation}
where $a_s=a_1 \eps^{1/3}$ is the radius of the equivalent sphere with the same volume as the ellipsoid, $\mu$ is the dynamic viscosity of the fluid and $\Kbpp$ represents the resistance tensor of an ellipsoidal particle in the co-moving frame \cite{fan95, zhang01}; see Supplemental Material for more details \cite{suppmat}. The resistance tensor $\Kbp$ is initially determined in the particle frame, and relates to $\Kbpp$ in the co-moving frame through the particle's rotation matrix $\mathbf{A}$, such that
\begin{equation}
\Kbpp = \mathbf{A^{-1}}\Kbp\mathbf{A} \text{.}
\end{equation}

In the case of a prolate ellipsoidal particle, e.g. invariant by rotation along its major axis, the resistance tensor reduces to a diagonal matrix, and for a 2-dimensional flow, we have
\begin{eqnarray*}
\Kbp = \frac{8}{3} \eps^{-1/3} \begin{pmatrix}
      \frac{\eps^2 - 1} { \frac{2(2\eps^2 - 1)}{\sqrt{\eps^2 - 1}} \ln ( \eps + \sqrt{\eps^2 - 1} )  - \eps} & 0 \\
0 & \frac{\eps^2 - 1} { \frac{   2\eps^2 - 3 }{\sqrt{\eps^2 - 1}} \ln ( \eps + \sqrt{\eps^2 - 1} )  + \eps}
\end{pmatrix}
\end{eqnarray*}
Substituting the drag force term applying to finite-size spheres in equation \eqref{MX3}, and noting $\Kbpp = 8/3 \eps ^{-1/3} \mathbf{\hat{\hat{k}}}$, we obtain
\begin{equation}\label{MX5}
\frac{d \vb}{dt} = \frac{D\ub}{Dt}  - \frac{16}{9}(\eps\St)^{-1} \;  \mathbf{\hat{\hat{k}}}  (\vb - \ub) \;\text{,}
\end{equation}
such that now, the equation of motion becomes coupled to the orientation of the particle given by equation \eqref{thetadot}. The drag force term's prefactor is proportional to $(\eps \St)^{-1}$, indicating that the relaxation time of the particle will now also depend on its aspect ratio; i.e. the more elongated the particle the longer the response time.
Similarly to Eq.\eqref{MX3}, drag forces can introduce perturbations and  also deviate elongated particles from fluid particle trajectories. Note that although the particles are passive, these perturbations allow the particles to cross flow streamlines and explore larger portions of the flow field, akin to active self-propelled particles \cite{babiano00, Torney2007, Khurana2011, Durham2011, Khurana2011}. However the dynamics exhibited by motile and inertial particles differ as they experience distinct forces. While prolate particles that can swim possess an intrinsic velocity aligned with the particle’s long axis at all times, passive particles on the other hand exhibit a distinct velocity from the fluid only when inertial forces are non negligible. These forces depend on both the particle’s orientation and the local flow geometry, such that, as opposed to motile particles with a fixed velocity, inertial particles intrinsic velocity can vary over time and point in different directions relative to the particle’s major axis.

\section{Materials and Methods}

\begin{figure}
\includegraphics[width= 17.2cm]{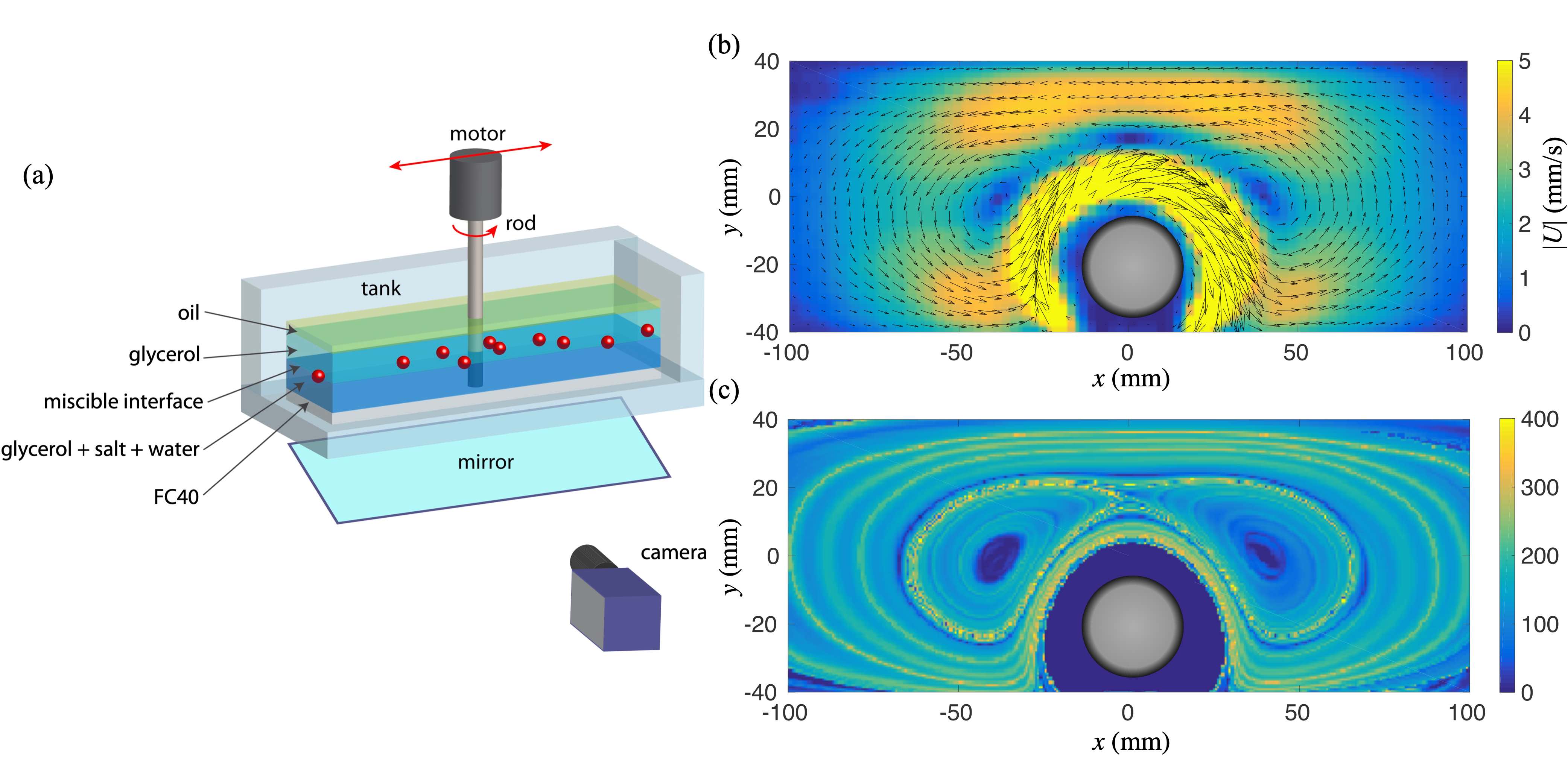}
\caption{(a) Experimental setup of the rotor-flow. One rotor attached to a traverse are respectively ensuring the rotation and translation of the rod. A mirror at 45 degrees is reflecting the recorded image of the interface from the bottom of the tank. The particles are positioned at a fixed height in a glycerol and salt density stratification. (b) Experimental flow velocity field determined from PIV measurements in the steady regime with a rotor speed at $200$ rpm and a fixed transversal position. The color code corresponds to the norm of the flow velocity. (c) Corresponding FTLE field computed for a time window $t = [0; 60]$ $\tau$ where $\tau = 0.3$ s is the rotor period.}
\label{PIV}
\end{figure}

The experimental setup consists in a rectangular container of dimensions $L \times W \times h = 100 \times 400 \times 200$ mm. To circumvent any surface tension effect \cite{vella2005}, the particles are trapped at the interface between two miscible fluids of different density. The tank is filled with two stratified layers of pure glycerol with density $\rho_1 = 1.26\pm0.001\text{ g.cm}^{-3}$ and a solution of glycerol and salt with $\rho_2 = 1.30 \pm 0.01\text{ g.cm}^{-3}$, as represented in Fig. \ref{PIV}(a). To prevent viscous instabilities at the interface, the viscosity of the layers are matched by adding $4\%$ water to the glycerol and salt solution. The particles have a density $\rho_p = 1.28\pm0.005$ and are stabilized by buoyancy forces on a quasi-2-dimensional plane, see Supplemental Material for pictures \cite{suppmat}. The bottom of the tank is filled with a 1 cm layer of low viscosity fluorinated oil to reduce friction on the lower wall, and a layer of vegetable oil is added on top to prevent water absorption by the glycerol over time. The particles motion is recorded with a 45 degree tilted mirror at the bottom the setup, and particle tracking velocimetry (PTV) is performed from digitized images with a 14 bit LaVision camera.

We use 3D-printing ABS filament of $1.75$ mm diameter with variable length to create anisotropic particles of the desired aspect ratio. To minimize particle-particle interactions, a small number of particles (32) is initially uniformly deposited and spread across the domain. For each particle's aspect ratio, five independent realizations were prepared manually by randomly redistributing the initial positions of the particles at the glycerol/glycerol-and-salt interface. Because of the stable stratification configuration and the high viscosity of the glycerol, the interface remained well define for several consecutive realizations of the experiment. The stratification is then renewed by changing the fluid mixture in the tank, and different particles are deposited at the interface.

The flow is created by stirring the fluid with a rotor attached to a longitudinal translation stage; see \cite{Filippi2020} for a similar experimental system description. The rotation is controlled with a Lexium NEMA 34 motion control motor with a $2$ cm diameter Plexiglas rod, and the translation of the rotor with a Parker lead-screw traverse attached on top of the tank. The resulting flow is quasi-2-dimensional in the stratified fluid layer, and similar to the rotor-oscillator flow topology described by Hackborn \cite{hackborn90, hackborn97, suppmat, weldon2008}. In the steady regime, when the position of the rotor is fixed, the rotation of the rod close to the wall creates a hyperbolic stagnation point in the center and two Moffatt eddies \cite{moffatt64} on each side of the rotor. Particle image velocimetry (PIV) is performed by seeding the upper glycerol layer with $30$ $\mu$m diameter hollow glass beads. A YAG laser $532$ nm is used to illuminate a sheet in the horizontal plane and visualize the flow near the plane containing the inertial particles. A typical flow velocity snapshot is displayed in Fig. \ref{PIV}(b). The inner region in the direct proximity of the rotor exhibits the highest velocities $\|u\| \gtrsim 10 \text{ mm.s}^{-1}$, and rapidly decays within a rod radius away from the rotating boundary. The two co-rotating Moffatt eddies are also visible, with an average flow speed in the outer region of $u \simeq 4 \pm 1 \text{ mm.s}^{-1}$ in their vicinity.

We perform in parallel numerical simulations solving the governing equations \eqref{MX3}, \eqref{thetadot} and \eqref{MX5} under identical conditions with the experiments. 
The size and the aspect ratio of the particles are set by choosing identical Stokes number and $\eps$ with the experimental particles. 
The particles initial positions are on a grid that matches our PIV field coverage, the initial orientation of the anisotropic particles $\theta_0$ are randomly chosen and the initial velocity is set slightly smaller than the fluid velocity at the particle's location $\vb(t=0) = 0.9\ub$.\\

\section{Numerical results}

\subsection{Focusing effect in the steady double gyre flow}

\begin{figure}
\includegraphics[width= 17.2cm]{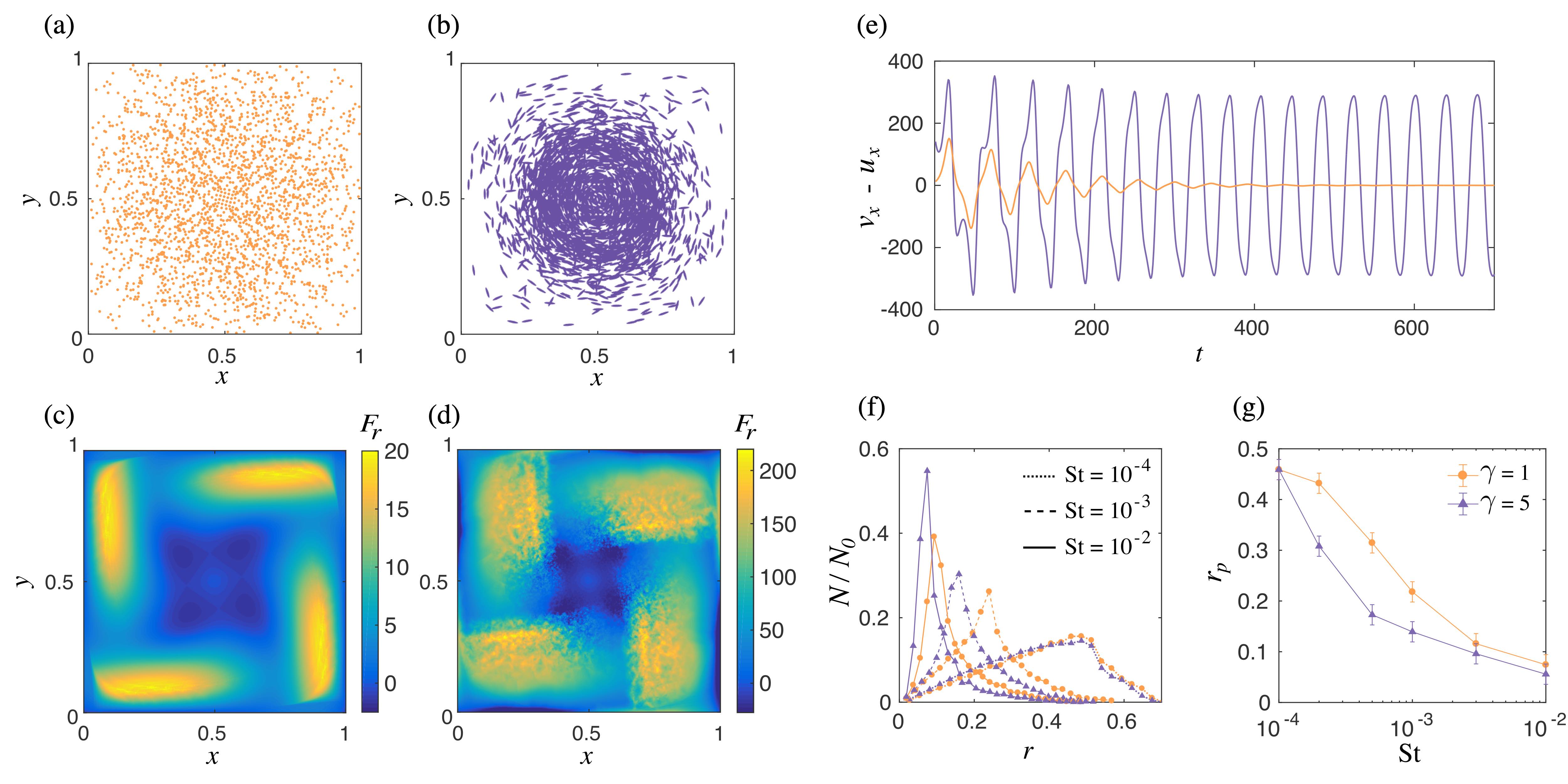}
\caption{Numerical particles positions in the steady regime for $t > 500$ in the steady vortex flow with $A=100$ and $\omega = 2 \pi / 10 $, for particles with $\St=5 \ 10^{-4}$ and initially at rest, (a) spheres and (b) ellipsoids with $\eps=5$. Corresponding radial component of the drag force experienced by the particles in the transient regime. Each of the color plots are normalized by the respective drag force prefactor $\St^{-1}$ and $(\gamma \St)^{-1}$ in Eq.~\eqref{MX3} and Eq.~\eqref{MX5}, (c) spheres and (d) ellipsoids with $\eps=5$. (e) Particle velocity $v_x$ and fluid velocity $u_x$ difference versus time at the particle's position for a spherical (orange) and ellipsoidal (purple) particle selected at the same initial position and with $\St=10^{-2}$. (f) Particles spatial distribution around the vortex center in the steady regime for 3 different St number, for spherical particles (orange circles), and ellipsoids with $\eps=5$ (purple triangles). The radial density profile is azimuthally averaged and normalized by the total number of  particles $N_0$. (g) Radial position of density profile maximum $r_p$ as a function of St.}
\label{fig3}
\end{figure}

We first numerically investigate the effect of particle size and shape on their trajectories over a large range of Stokes number $ 10^{-4} \leqslant \St \leqslant 10^{-1} $, and aspect ratio $1\leqslant \eps \leqslant 10 $. We consider a double gyre flow with periodic boundary conditions defined by the streamfunction
\begin{equation}\label{vortex_flow}
\psi(x,y,t) = A \sin [(\pi f(x,t)] \sin(\pi y) \text{,}
\end{equation}
where $A$ sets the flow amplitude, $f(x,t)= x + B(x^2  - 2) \sin \omega t$, and $B$ sets the oscillation amplitude. In the steady regime $B=0$, and the streamfunction reduces to $\psi(x,y) = A \sin (\pi x) \sin(\pi y)$, leading to a simple periodic array of counter-rotating vortices.

The resulting spatial distributions of neutrally buoyant particles with $\St=5 \ 10^{-4}$ and different shapes around a single vortex are shown in Figs. \ref{fig3}(a) and \ref{fig3}(b). The streamfunction amplitude is set to $A=100$, such that all the particles converge rapidly to a steady spatial distribution after a short transient regime and ellipsoidal particles orientation aligns with the local flow streamlines. Figures \ref{fig3}(c) and \ref{fig3}(d) show the corresponding radial component of the drag forces during the early transient dynamics. For both types of particles, a negative radial component is displayed near the center of the vortex, and indicates attractive forces toward this region of the flow. Drag forces are systematically larger for the ellipsoidal particles, leading to stronger deviations in average than for spherical particles. As can be seen in Fig. \ref{fig3}(e), even for the largest simulated Stokes number $\St=10^{-2}$, spherical particles velocity rapidly relaxes toward the fluid flow velocity while elongated particles velocity doesn't converge to the flow velocity. Anisotropic particles systematically present a stronger focusing effect, converging toward smaller orbit radii than the spherical particles at equal $\St$ number. To illustrate this effect, we repeat the same simulation and vary the St number by two order of magnitude. Figure \ref{fig3}(f) shows the resulting radial distribution of particles in the steady regime. As the St number increases, the particles spatial distribution becomes more peaked and moves closer to the vortex center. As can be seen in Fig. \ref{fig3}(g), the distance of the particle density peak to the vortex center $r_p$ decreases faster for ellipsoidal particles as a function of the St number before becoming similar for both types of particles when $\St \gtrsim 3\ 10^{-3}$.

Even at small $\St$ number $\St \approx 10^{-4}$, when inertial effects are expected to be negligible for spherical particles, anisotropic particles exhibit a more complex dynamics with a stronger attraction to vortex cores.

\subsection{Particles dispersion in time-varying flows}

\begin{figure}
\includegraphics[width= 18cm]{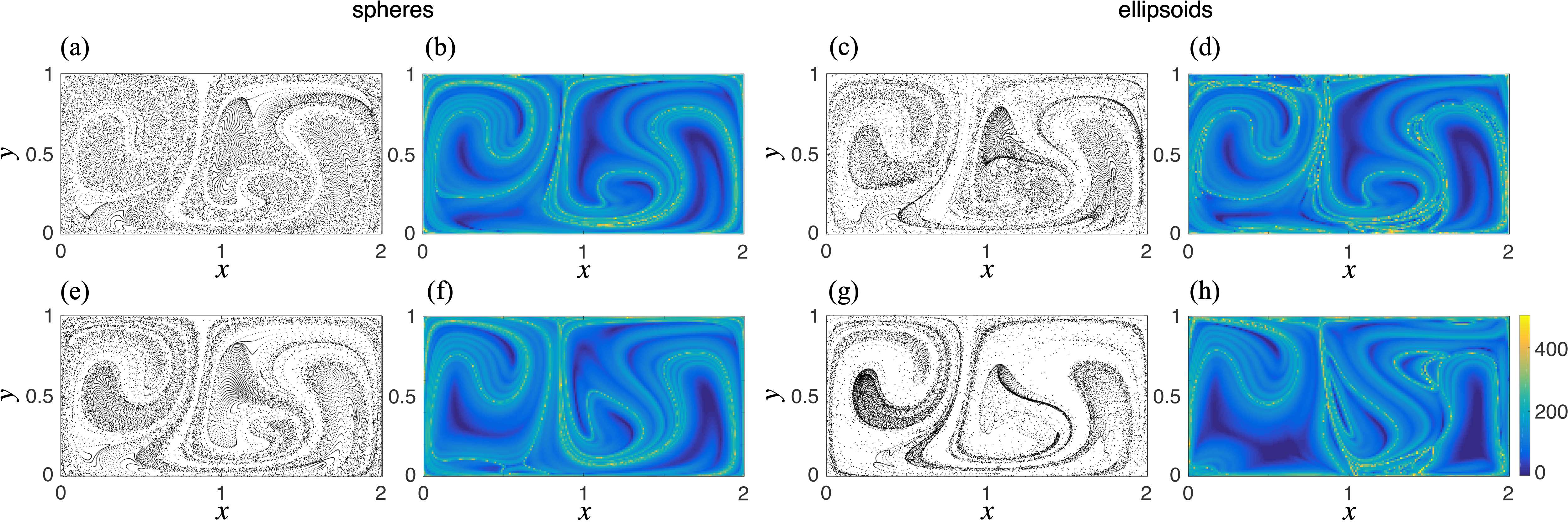}
\caption{Inertial particles positions and corresponding backward FTLE fields in simulations for spherical particles, left column, and ellipsoidal particles with $\eps = 5$, right column, after $t=20$ time steps in a double gyre flow with $A=100$, $e=0.25$, and $\omega = 2 \pi A$. (a)-(d) $\St = 10^{-3}$ and particles initial velocity $\vb_0 = 0.9 \ub$, (e)-(h) $\St = 5\ 10^{-4}$ and $\vb_0 = 0$.}
\label{DG_time}
\end{figure}

We now use the full time-dependent double gyre flow, and set $e = 0.25$. In time varying flows, attractive and repulsive structures are only defined over a finite-time period. In this context, Lagrangian coherent structures (LCS) provide a useful framework to understand the transport of material in complex, unsteady flow fields \cite{peacock2013}. To identify these regions, we first compute the finite time Lyapunov exponent (FTLE) field corresponding to our flow and the inertial particles trajectories for a given time period \cite{sudharsan2016, suppmat, pierrehumbert93, haller01}.
Regions with a low FTLE value are indicative of attractive properties; i.e. initially neighboring particles remain close to each other over a given time window, while regions with a high value correspond to repulsive regions of the flow, where the spatial separation over the same time window between neighboring fluid particles is maximum \cite{haller2015}.

Figure \ref{DG_time} displays spatial distributions and corresponding FTLE fields for inertial particles with two different initial velocity and $\St$ number. In all configurations, after $t=2/A$, where $A$ is the gyre rotation frequency, the ellipsoidal particles distribution systematically present a more inhomogeneous spacial density variation than the spherical ones. At $\St = 5\ 10^{-4} $, the corresponding spherical particles FTLE field is similar to the flow FTLE field, see Supplemental Material \cite{suppmat}. The ellipsoidal particles FTLE fields are altered and exhibit additional structures located in regions with stronger shear. This suggests that in time varying flows, ellipsoidal particles more complex dynamics can lead to a departure from passive transport even at very small St numbers.

\section{Experimental results}

\subsection{Focusing effect in the rotor-flow}

\begin{figure}
\centering
\includegraphics[width= 17.2cm]{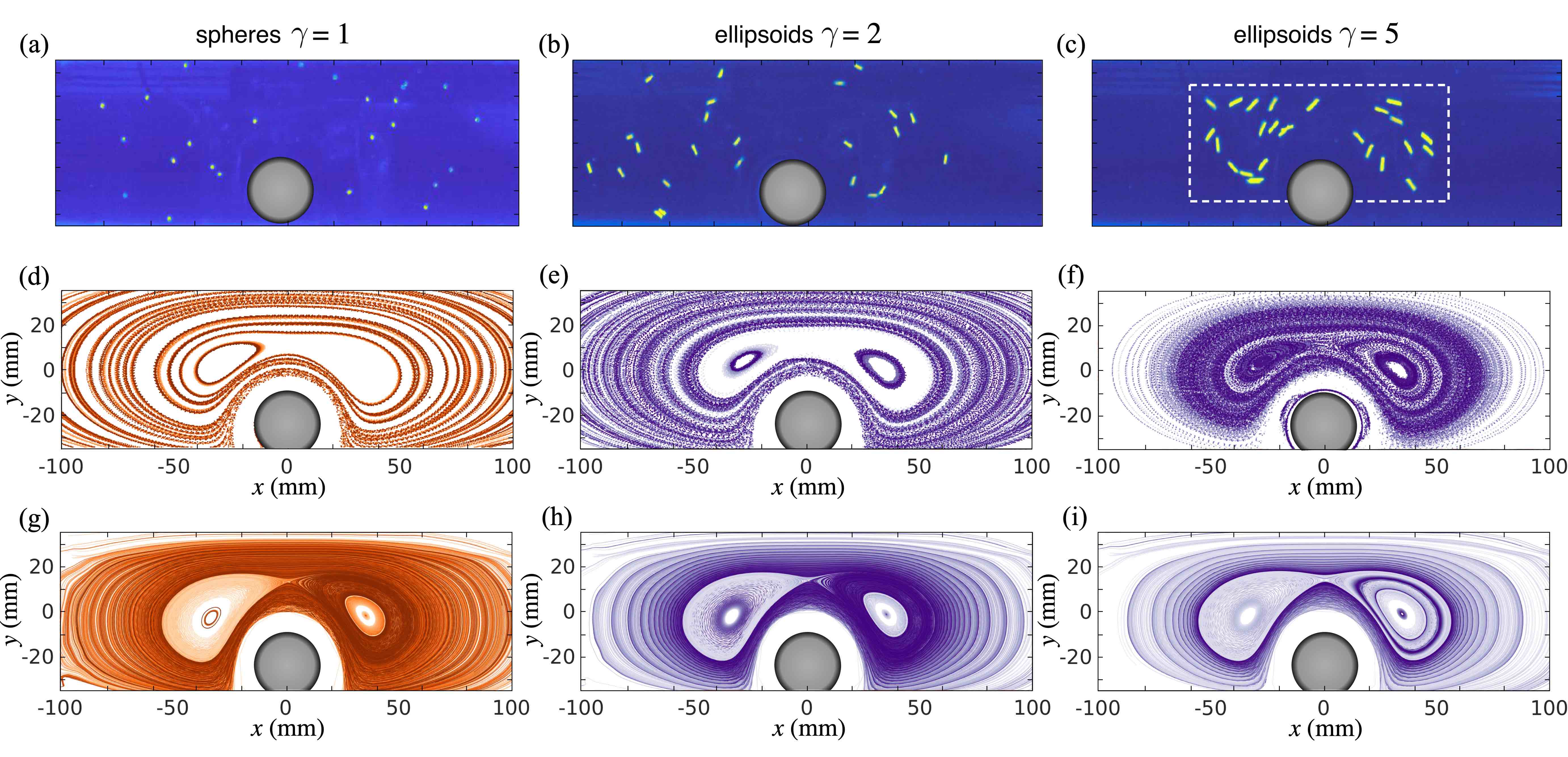}
\caption{(a)-(c) experimental particles final positions at $t = 12000$ $\tau$, (d)-(f) particles positions over a time window $t = [5000; 9200]$ $\tau$ in the corresponding experiments, (g)-(i) numerical particles positions advected in the experimental flow field shown in Fig. \ref{PIV}(b) for the entire duration of the simulation equivalent to $t = 6700$ $\tau$. Spherical (orange) and ellipsoidal (purple) particles, color intensity indicates time.}
\label{ptv}
\end{figure}

To test our findings, we perform a series of experiments and compare the resulting trajectories with our model by computing the numerical particles dynamics in the same, experimentally measured, flow field. The flow has the geometry of a rotor-flow in the steady regime, and a snapshot of the experimental velocity field with the corresponding FTLE field is shown in Figs. \ref{PIV}(b) and (c). We impose a constant rotor speed of $200$ rpm for all three different particle aspect ratio $\eps =$ 1, 2 and 5. In this configuration, the Reynolds number typically ranges between $0.1 \leq \re \leq 10$ in the flow central region, away from the direct vicinity of the rotor indicated by the gray disk in Figs. \ref{PIV} and \ref{ptv}. This leads to relatively small particle Stokes number $\St \approx 10^{-4} - 10^{-3}$, such that, in the absence of anisotropic effects, we expect the particles with $\eps = 1$ to behave closer to tracer particles in our experiments.

The particles are initially distributed uniformly across the experimental domain with no preferential alignment across the quasi-2-dimensional plane defined by the two-layer stratification, as described in the Method section \cite{suppmat}.
The particles remain in the same plane for the entire duration of the experiments, and the principal axis of the elongated particles doesn't display off-plane motions except near the rotor wall.
Immediately after the flow is started, particles with $\eps > 1$ change their orientation and exhibit a preferential alignment of their principal axis with the local flow direction, as shown in the supplemental movie 1 \cite{suppmat}. This suggests that the orientational behavior has a negligible response time to the changes in the flow field, in good agreement with equation \eqref{thetadot}, and previously reported findings for flows at higher Reynolds number \cite{parsa11,ni14}.

After a time $t \gtrsim 4000$ $\tau$, where $\tau=0.3$ s corresponds to the rotor rotation time period, the particles spatial distribution approaches a stationary regime which strongly depends on their aspect ratio, as shown in Figs. \ref{ptv}(a)-(c) and the supplemental movies 2-4 \cite{suppmat}.
The final position of the particles with aspect ratio $\eps=1$ do not exhibit any preferential concentration, and remain uniformly spread across the domain, while particles with $\eps=2$ and $\eps=5$ display a migration from the outer regions of the flow into the vortex cores and their vicinity. Remarkably, for $\eps=5$, all the particles are focused in the central region. Note that for $\eps=2$ and $\eps=5$, some particles can be attracted to the secondary vortices located near the lateral edges of the tank not covered in our field of views in Figs. \ref{PIV} and \ref{ptv}.

In order to quantify this behavior, we determine the individual particles trajectories using particle tracking velocimetry (PTV), and monitor their spatial distribution over time. Figures \ref{ptv}(d)-(f) display all the particles positions as a function of time for the entire duration of the experiment shown in Figs \ref{ptv}(a)-(c) respectively.  
As can be seen in Fig. \ref{ptv}(e) and \ref{ptv}(f), some of the elongated particles' final trajectories form closed orbits inside the vortices. While for $\eps=2$, the particles remain more evenly distributed outside the vortices, particles with $\eps=5$ exhibit a strong concentration in the direct vicinity of the vortices.

Figure \ref{box} displays the measured particle density variation between $0 \leq t \leq 12000$ $\tau$ in the central region of the flow delimited by the white dashed lines in Fig. \ref{ptv}(c). During a first transient regime, the spatial density of the elongated particles increases linearly over time and exhibit an aspect ratio dependent migration rate, with the largest rate for $\eps=5$. After $t= t^*$, the migration slows down and the density of particles in the central region of the flow converges toward a constant density over time. Interestingly, while both the migration rate and the final particle concentration depend on $\gamma$, the saturation time scale $t^* \simeq 4000$ $\tau$ seems similar for particles with aspect ratio $\eps > 1$. Particles with $\eps=1$ present a negligible focusing effect, and their spatial distribution remains nearly unchanged over time.

\subsection{Comparison with the numerical model}

We perform in parallel numerical simulations using the experimentally measured flow field shown in Fig. \ref{PIV}(b). The particles $\St$ number and aspect ratio are now chosen identical to the experimental particles. The flow field is first nondimensionalized as described in Supplemental Material \cite{suppmat}. We set $x^* = x/L$ with $L = 100$ mm corresponding to the width of the tank, and $u^*=u/U$, where $U=L/T$ is the characteristic flow velocity. To estimate $U$ from the experiments, we select the velocity that gives the best match between the numerical and experimental particles velocities located in identical regions of the flow.
The best agreement with the experimental particles velocities is obtained for $U = 5 \pm 2$ mm.s$^{-1}$, in good agreement with the largest flow structures average velocity in Fig. \ref{PIV}(b).

\begin{figure}
\includegraphics[width= 8.6cm]{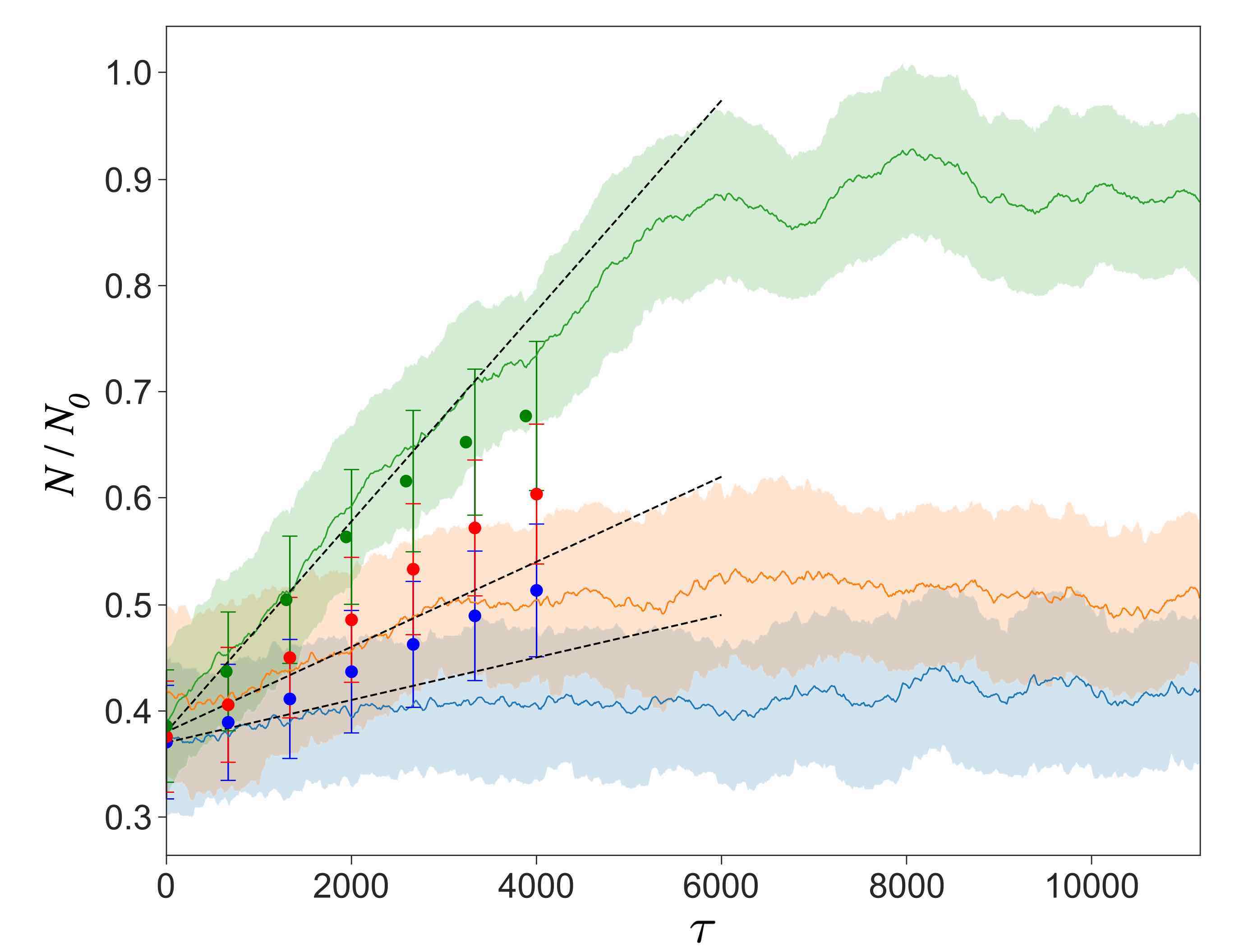}
\caption{Spatial particle density variation as a function of time in the central region of the flow indicated by the white dashed lines in Fig. \ref{ptv}c). Lines with shaded region, experimental average value and corresponding error. Circles, numerical average value with $\St = 5 \ 10^{-4}$ in the experimentally measured flow field.  Blue : $\eps = 1$, orange: $\eps = 2$, and green: $\eps = 5$. Dashed lines, linear fit of the experimental data in the transient regime for $t<t^*$.}
\label{box}
\end{figure}

The resulting numerical trajectories are shown in Figures \ref{ptv}(g)-(i), and qualitatively agree with the experimental trajectories spatial features. They display a similar convergence toward the vicinity of the vortices for $\eps >1$, although a larger asymmetry between the left and right attractive regions within the vortices is visible, and only the right vortex traps the most elongated particles on a distinct inner orbit. The migration rate of the numerical particles are shown in Fig. \ref{box}, and display a similar trend with the experimental data in the transient regime. After $t \gtrsim 3000$ $\tau$, the predicted aggregation rates start differing from the experimental curves and lead to a less contrasted particle spatial distribution as a function of $\eps$.

Comparing the shape of high particle concentration regions in our experiments and simulations with the rotor-flow FTLE field shown in Fig. \ref{PIV}(c), we can identify which flow structures attract the most the elongated particles. The flow FTLE field displays several ridges: two minima within the vortex cores corresponding to the two ellipsoidal regions where elongated particles are trapped in the experiments, and a third structure formed by the minima encircling the two vortices. Both experimental and numerical particles with the largest aspect ratio $\eps = 5$ are specifically attracted to this third region and display the large closed orbit visible in Figs. \ref{ptv}(f) and \ref{ptv}(i).\\

\section{Discussion}

In our experiments, the observed migration toward vortices for elongated particles with a small Stokes number $10^{-4} \lesssim \St \lesssim 10^{-3}$, suggests that even weak deviations from passive trajectories can build up over time, leading to distinct spatial organization over long time periods. These perturbations are not sufficient to significantly alter particles trajectories with an aspect ratio $\eps = 1$, and particles with $\eps = 2$  exhibit a considerably weaker focusing effect than those with $\eps = 5$.
A direct comparison of these observations with our numerical simulations shows that our minimal model qualitatively captures this behavior, although the particles focusing effect exhibits a smaller dependence to the particle aspect ratio $\eps$ than in the experiments. This difference could be attributed to second order orientation dependent terms, such as the added mass \cite{gallily79-2, koch89, mortensen08, zhang01, dibenedetto2018}, that were neglected here.

These results suggest that anisotropic particles respond to the flow with a larger lag. In addition to the orientation dependent nature of the drag forces in equation \eqref{MX5}, $\St^\prime = \eps\St$ could be interpreted as an ``effective''  Stokes number that is proportional to $\eps$. For weak inertial effects $\St \ll 1$, the difference between $\St$ and $\St^\prime$ on the particles trajectories becomes only visible at large time scales, when small deviations have accumulated over longer time periods.

This effect could potentially enhance the migration rate of irregularly shaped objects such as microplastic in the ocean toward gyres of marine debris \cite{chubarenko2016, dibenedetto2018}, and influence the garbage patches formation dynamics \cite{broday1997, sebille12}.
On the opposite length and time scales, in biology this interplay between the shape of the particle and the flow field geometry could influence the selective capture of microorganisms by larger animals. For instance, flows created by the beating of cilia can generate a local flow that attract only microorganisms of an optimal size \cite{ding2015}. Identical mechanisms could potentially select for an optimum shape, such as the aspect ratio of a given bacteria.

In conclusion, even neutrally buoyant particles with small particle Stokes number $\St \simeq 10^{-4}$ exhibit trajectories that deviate from the fluid flow when they possess an anisotropic shape. Both experimental and numerical particles exhibit a migration toward regions in the vicinity of vortices. 
The orientational dependence of the drag force term in our minimal model describes qualitatively the particles dynamics, suggesting that this mechanisms could be the dominant effect for the separation of isotropic and anisotropic particle trajectories in cellular flows.
In addition, the effect of additional orientation dependent terms and experiments with unsteady flows should be investigated in future work.

\section{Acknowledgments}
The authors would like to thank A.W. Murray and E. Climent for stimulating discussions, G. Haller, T. Witten and E. Kanso for their thought provoking comments. S. A. was supported by the French government DGA/DS - Mission pour la Recherche et l'Innovation Scientifique fellowship during her visit at MIT. T. P. acknowledges funding from ONR Grant N000141812762.

\bibliography{ILCS}

\end{document}